\documentclass[twocolumn,aps,prl,superscriptaddress,amsfonts]{revtex4-1}

\usepackage{amsmath}
\usepackage{graphicx}
\usepackage{bm}
\usepackage{array}
\usepackage{dcolumn}
\usepackage{hepparticles}
\usepackage{heppennames}

\begin{document}

\DeclareRobustCommand{\Hbar}{\HepAntiParticle{H}{}{}\xspace}
\DeclareRobustCommand{\H}{\HepParticle{H}{}{}\xspace} \newcommand{\pbar}{\APproton}

\DeclareRobustCommand{\pbar}{\HepAntiParticle{p}{}{}\xspace}
\DeclareRobustCommand{\p}{\HepParticle{p}{}{}\xspace}

\DeclareRobustCommand{\pos}{\HepParticle{e}{}{+}\xspace}
\DeclareRobustCommand{\e}{\HepParticle{e}{}{}\xspace}

\DeclareRobustCommand{\mup}{$\mu_{p}${}{}\xspace}
\DeclareRobustCommand{\muN}{$\mu_N${}{}\xspace}
\DeclareRobustCommand{\mupbar}{$\mu_{\pbar}${}{}\xspace}

\newcommand{\zhat}{\bf \hat{z}}
\newcommand{\yhat}{\bf \hat{y}}
\newcommand{\xhat}{\bf \hat{x}}

\newcommand{\NowYork}{\altaffiliation{Current address: York Univ., Dept.\ of Phys.\ and Astr., Toronto, Ontario M3J 1P3, Canada}}

\newcommand{\Harvard}{\affiliation{Dept.\ of Physics, Harvard University, Cambridge, MA 02138}}
\newcommand{\Juelich}{\affiliation{IKP, Forschungszentrum J\"{u}lich GmbH, 52425 J\"{u}lich, Germany}}
\newcommand{\JuelichZat}{\affiliation{ZAT, Forschungszentrum J\"{u}lich GmbH, 52425 J\"{u}lich, Germany}}
\newcommand{\York}{\affiliation{York University, Department of Physics and Astronomy, Toronto, Ontario M3J 1P3, Canada}}
\newcommand{\Garching}{\affiliation{Max-Planck-Institut f\"{u}r Quantenoptik, Hans-Kopfermann-Strasse 1, 85748 Garching, Germany}}
\newcommand{\Munich}{\affiliation{Ludwig-Maximilians-Universit\"{a}t M\"{u}nchen, Schellingstrasse 4/III, 80799 M\"{u}nchen, Germany}}
\newcommand{\Mainz}{\affiliation{Institut f\"{u}r Physik, Johannes Gutenberg Universit\"{a}t and Helmholtz Institut Mainz, D-55099 Mainz, Germany}}
\newcommand{\MainzWalter}{\affiliation{Institut f\"ur Physik, Johannes Gutenberg Universit\"at Mainz, D-5509, Mainz, Germany}}

\title{One-Particle Measurement of the \pbar Magnetic Moment}

\author{J.\ DiSciacca}
\author{M.\ Marshall}
\author{K.\ Marable}
\author{G.\ Gabrielse}\email[Spokesperson: ]{gabrielse@physics.harvard.edu}
\author{S.\ Ettenauer}
\author{E.\ Tardiff}
\author{R.\ Kalra}
\Harvard

\author{D.W.\ Fitzakerley}
\author{M.C.\ George}
\author{E.A.\ Hessels}
\author{C.H.\ Storry}
\author{M.\ Weel}
\York

\author{D.\ Grzonka}
\Juelich
\author{W.\ Oelert}
\Juelich
\MainzWalter
\author{T.\ Sefzick}
\Juelich

\collaboration{ATRAP Collaboration}\noaffiliation

\date{Submitted to PRL on 22 Jan.\ 2013}

\begin{abstract} 
For the first time a single trapped \pbar is used to measure the \pbar magnetic moment ${\bm\mu}_{\pbar}$.  The moment ${\bm\mu}_{\pbar} = \mu_{\pbar} {\bm S}/(\hbar/2)$ is given in terms of its spin ${\bm S}$ and the nuclear magneton (\muN) by $\mu_{\pbar}/\mu_N = -2.792\,845 \pm 0.000\,012$.  The 4.4 parts per million (ppm) uncertainty is 680 times smaller than previously realized. Comparing to the proton moment  measured using the same method and trap electrodes gives  $\mu_{\pbar}/\mu_p = -1.000\,000 \pm 0.000\,005$ to 5 ppm, for a proton moment ${\bm{\mu}}_{p} = \mu_{p} {\bm S}/(\hbar/2)$, consistent with the prediction of the CPT theorem.  
\end{abstract}

\pacs{37.10.Ty, 13.40.Em, 14.20.Dh}

\maketitle

Measurements of the properties of particles and antiparticles are intriguing in part because the fundamental cause of the asymmetry between matter and antimatter in the universe has yet to be discovered.  Within the standard model of particle physics, the results of particle-antiparticle comparisons are predicted  by a CPT theorem \cite{Luders57} that pertains because systems are described by a local, Lorentz-invariant, quantum field theory (QFT).  Whether the theorem applies universally is open to question, especially since gravitational interactions have so far eluded a QFT description. It is thus important to precisely test predictions of the CPT theorem, one example of which is that \pbar and \p magnetic moments have opposite signs and the same magnitude.  Testing this prediction may eventually produce a second precise CPT test with a baryon and antibaryon, of comparable precision to the \pbar and \p charge-to-mass ratio comparison \cite{FinalPbarMass}.

\begin{figure}[htbp!]
\includegraphics*[width=\columnwidth]{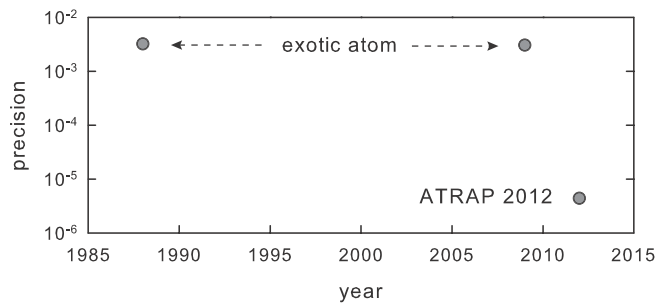}
\caption{Uncertainties in measurements of the \pbar magnetic moment measured in nuclear magnetons, $\mu_{\pbar}/\mu_N$.}
\label{fig:MagneticMomentHistory}
\end{figure}

The \pbar magnetic moment was previously deduced only from measured transition energies in exotic atoms in which a \pbar orbits a nucleus as a ``heavy electron.''  Measurements 24 and 4 years ago \cite{1988PbarMoment,2009PbarMoment} both reached a 3000 ppm precision (Fig.~\ref{fig:MagneticMomentHistory}). Meanwhile, single particle methods were used to measure other magnetic moments to a much higher precision. For example, the most precisely measured property of an elementary particle is the electron magnetic moment measured with one electron \cite{HarvardMagneticMoment2008}.

This Letter reports the first single-particle measurement of the \pbar magnetic moment, a 4.4 ppm determination that is 680 times more precise than realized with exotic atoms (Fig.~\ref{fig:MagneticMomentHistory}).  The methods and apparatus were initially demonstrated in a one-proton measurement of \mup \cite{ProtonMagneticMomentWithCorrection}, following the realization of feedback cooling and a self-excited oscillator with one proton \cite{OneProtonSelfExcitedOscillator}. We profited from a parallel exploration of proton spin flips \cite{MainzSpinFlips} and a measurement of \mup \cite{MainzProtonMoment2012} that followed ours.  

The cyclotron and spin frequencies ($f_c$ and $f_s$), measured for a single \pbar suspended in a magnetic field, determine the \pbar moment in nuclear magnetons, 
\begin{equation}
\frac{\mu_{\pbar}}{\mu_N} \equiv \frac{g_{\pbar}}{2} \frac{q_{\pbar}/m_{\pbar}}{q_p/m_p}\approx -\frac{g_{\pbar}}{2} = -\frac{f_s}{f_c}, 
\label{eq:ProtonMagneticMoment}
\end{equation}
where $g_{\pbar}$ is the \pbar $g$-factor.  The ratio of \pbar and \p charge-to-mass ratios enters because the nuclear magneton \muN is defined in terms of the proton charge and mass.  This ratio was measured to be -1 to 0.0001 ppm using a \pbar simultaneously trapped with a $H^-$ ion \cite{FinalPbarMass} so the approximation in Eq.~\ref{eq:ProtonMagneticMoment} is more than adequate for our precision.

\begin{figure}[htbp!]
\centering
\includegraphics*[width=\columnwidth]{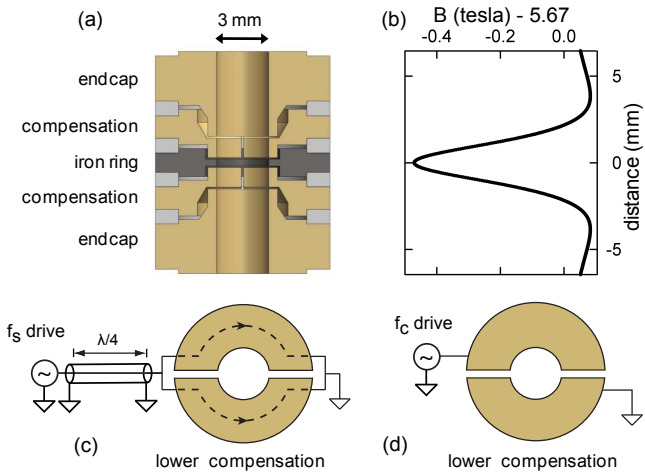}
\caption{(a) Electrodes of the analysis trap (cutaway side view) are copper with an iron ring. (b) The iron ring significantly alters B on axis. (c) Top view of the paths of the oscillating current for the spin flip drive.  (d) An oscillating electric field (top view) drives \pbar cyclotron motion. }
\label{fig:PbarMeasurementTrap}
\end{figure}

The \pbar magnetic moment is measured within the ``analysis trap'' electrodes (Fig.~\ref{fig:PbarMeasurementTrap}) used to measure the proton magnetic moment \cite{ProtonMagneticMomentWithCorrection}.  
The stacked rings are made of OFE copper or iron, with a 3 mm inner diameter and an evaporated gold layer.  The electrodes and surrounding vacuum container are cooled to 4.2 K by a thermal connection to liquid helium.  Cryopumping of the closed system made the vacuum better than $5 \times 10^{-17}$ Torr in a similar system \cite{PbarMass}, so collisions are unimportant.  Appropriate potentials applied to electrodes with a carefully chosen relative geometry \cite{OpenTrap} make a very good electrostatic quadrupole near the trap center with open access to the trap interior from either end.

After the proton measurement \cite{ProtonMagneticMomentWithCorrection} was completed, the apparatus was modified and moved from Harvard to CERN.  The neighboring electrodes and vacuum enclosure (not pictured in Fig.~\ref{fig:PbarMeasurementTrap}) were modified to allow 5 MeV \pbar from CERN's Antiproton Decelerator (AD) to enter the vacuum enclosure through a thin Ti window and to be captured and electron-cooled in the neighboring electrodes.  The cooling electrons are ejected by reducing the trap potential long enough that light electrons escape while heavier \pbar do not. These methods, now used for all low energy \pbar and \Hbar experiments, are reviewed in \cite{PbarReview}.

Once the \pbar is centered in the analysis trap, in a 5 tesla vertical magnetic field $\mathbf{B} =B \zhat$ tesla, the circular cyclotron motion of a trapped \pbar is perpendicular to {\bf B} with a frequency $f_+ = 79.152$ MHz
slightly shifted from $f_c$  by the electrostatic potential.  The \pbar also oscillates parallel to {\bf B} at about $f_z =
\, 920$ kHz.    The third motion is a circular magnetron motion, also perpendicular to
{\bf B}, at the much lower frequency $f_- = 5.32$ kHz.  The spin precession frequency is $f_s=221.075$ MHz.

Driving spin flips requires a magnetic field perpendicular to {\bf B} that oscillates at approximately $f_s$.  This field is generated by currents (increased compared to \cite{ProtonMagneticMomentWithCorrection} by a transmission line transformer) sent through halves of a compensation electrode (Fig.~\ref{fig:PbarMeasurementTrap}c).  Driving  cyclotron transitions requires an electric field perpendicular to {\bf B} that oscillates at approximately $f_+$.  This field is generated by potentials applied across halves of a compensation electrode (Fig.~\ref{fig:PbarMeasurementTrap}d).

Much of the challenge of the measurement arises from the small size of a nuclear magnetic moment.  
Unlike the electron moment, which scales naturally as a Bohr magneton ($\mu_B$), the nuclear moments scale as the much smaller nuclear magneton $\mu_N,$ with $\mu_N/\mu_B = m_e/m_p \sim 1/2000$.  Shifts in $f_z$ reveal changes in the cyclotron, spin and magnetron quantum numbers $n$, $m_s$ and $\ell$ \cite{Review},
\begin{equation} \frac{\Delta f_z}{f_z} = \frac{\hbar
 \beta_2}{4 \pi m_p |B| f_- } \left( n + \frac{1}{2} + \frac{g_p m_s}{2} + \frac{f_-}{f_+} (\ell + \frac{1}{2})
\right).  \label{eq:FrequencyShift}
\end{equation}
The shifts (50 and 130 mHz per cyclotron quantum and spin flip) arise from 
a saturated iron ring (Fig.~\ref{fig:PbarMeasurementTrap}a) that adds (to {\bf B})
a magnetic bottle gradient (at trap center), 
\begin{equation}
\Delta {\bf B} = \beta_2 [(z^2-\rho^2/2){\bf \hat{z}} - z\rho \boldsymbol{\hat{\rho}}]. 
\end{equation}
The effective $f_z$ shifts because the electrostatic axial oscillator Hamiltonian going as $f_z^2 z^2$ acquires $\mu z^2$ from the interaction of cyclotron, magnetron and spin moments, $\mu \zhat$, with $\Delta {\bf B}$. 
The bottle strength, $\beta_2 = 2.9 \times 10^5$ T/m$^2$, is 190 times that used to detect electron spin flips \cite{HarvardMagneticMoment2008} to compensate partially for the small $\mu_N$.

The \pbar are transferred between the analysis trap and an adjacent coaxial trap  (not in Fig.~\ref{fig:PbarMeasurementTrap}) by slowly varying the applied electrode voltages to make the axial potential well move adiabatically between the two trap centers. In the adjacent trap the \pbar cyclotron motion induces currents in and comes to thermal equilibrium with an attached damping circuit cooled to near 4 K.  The cooled \pbar is transferred back to the analysis trap and a measured shift $\Delta f_z < 100 $Hz is required to ensure a cyclotron radius below 0.7 $\mu m$ (a bit larger than was possible with more time in \cite{ProtonMagneticMomentWithCorrection}) before measuring $f_s$.  For larger shifts, the \pbar is returned to the precision trap for cyclotron damping as needed until a low cyclotron energy is selected.

Two methods are used to measure  the $\Delta f_z$ of Eq.~\ref{eq:FrequencyShift} in the analysis trap, though the choice of which method to use in which context is more historical than necessary at the current precision.  The first (used to detect cyclotron transitions with the weakest possible driving force) takes $\Delta f_z$ to be the shift of the frequency at which Johnson noise in a detection circuit is canceled by the signal from the \pbar axial motion that it drives \cite{DehmeltWalls1968}.  The second (used to detect spin flips) takes $\Delta f_z$ to be the shift of the frequency of a self-excited oscillator (SEO) \cite{OneProtonSelfExcitedOscillator}.  The SEO oscillation arises when amplified signal from the \pbar axial oscillation is fed back to drive the \pbar into a steady-state. 

\begin{figure}[htbp!]
\includegraphics*[width=\columnwidth]{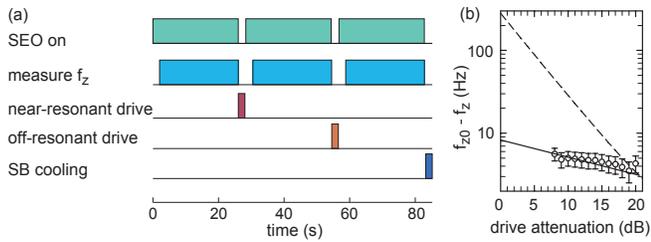}
\caption{(a) Spin measurement cycle. (b) The power shift in $f_z$ due to the spin flip drive (points) is lower than was observed for the proton measurement with no transmission line transformer (dashes).}
\label{fig:MeasurementSequence}
\end{figure}

The measurement cycle in Fig.~\ref{fig:MeasurementSequence} is used to find spin resonance. After the SEO stabilizes for 2 s, its frequency average over 24 seconds is $f_1$.  With the SEO off, a nearly-resonant spin flip drive at frequency $f_d$ is applied for 2 s.  After the SEO is back on for 2 s, the average frequency $f_2$ is measured to determine the deviation $\Delta = f_2 - f_1$.  A spin drive detuned 100 kHz from resonance is next applied with the SEO off -- detuned rather than off to control for possible secondary frequency shifts due to the drive. The average $f_3$ is then measured and compared to $f_2$.  The cycle concludes with 2 s of sideband cooling to prevent magnetron radius growth \cite{OneProtonSelfExcitedOscillator}.   

Repeating the measurement cycle yields a sequence of deviations $\Delta_i$ that can be characterized by an Allen variance $\sigma^2 = \sum_{i=1}^{N} \Delta_i^2/(2N)$ (often used to describe the stability of frequency sources).
Even when no spin drive is applied, jitter in the axial frequency (not well understood \cite{OneProtonSelfExcitedOscillator}) gives the Allan variance a nonzero value $\sigma_0^2$ comparable to the deviation caused by a spin flip, .  This jitter increases with cyclotron radius so $\sigma_0^2$ is reduced by selecting a \pbar with small cyclotron radius (as described earlier).   When a drive at frequency $f_d \approx f_s$ induces spin flips, the Allan variance increases slightly to $\sigma_f^2=\sigma_0^2 + \sigma^2$.

Both the spin and cyclotron resonances are expected to show no excitation until the drive frequency increases above a sharp threshold  \cite{BrownLineshape,Review}.  The driving force has no effect below a resonance frequency ($f_+$ or $f_s$ here).  The transition rate between quantum states and the resulting broadening increases abruptly to its maximum at the resonant frequency. Above this threshold there is a distribution of  cyclotron or spin frequencies at which these motions can be driven.  These correspond to the distribution of $B$ sampled by the thermal axial motion of the \pbar (in thermal equilibrium with the axial detection circuit) within the magnetic bottle gradient.

The spin and cyclotron motions are not damped in the analysis trap so natural linewidth does not broaden the sharp threshold edges.   The superconducting solenoid produces a stable B that does not significantly broaden the edge.  A small broadening arising because sideband cooling (of magnetron motion coupled to  axial motion)
selects different values from a distribution of magnetron radii (explored in detail in \cite{OneProtonSelfExcitedOscillator}) is added as ``magnetron broadening'' uncertainty in Table \ref{table:Uncertainties}.

\begin{figure}[htbp!]
\includegraphics*[width=\columnwidth]{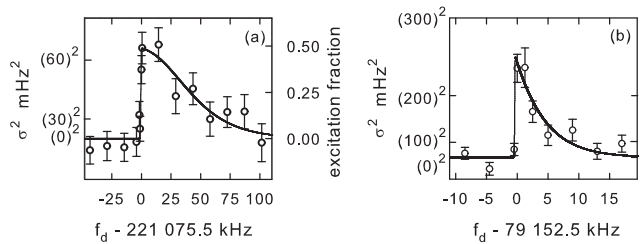}
\caption{(a) The spin line. (b)The cyclotron line.}
\label{fig:CyclotronAndSpinFlipLineshape}
\end{figure}

For each drive frequency in Fig.~\ref{fig:CyclotronAndSpinFlipLineshape}a the cycle in Fig.~\ref{fig:MeasurementSequence} is repeated for 24 to 48 hours.  The Allan deviation  $\sigma_f$ for the sequence of deviations $\Delta_f=f_2-f_1$ represents the effect of fluctuations when a near-resonant spin drive is applied.  The Allan deviation $\sigma_0$ for the sequence of deviations $\Delta_0=f_3-f_2$ represents fluctuations when no near-resonant drive is applied. The spin lineshape in Fig.~\ref{fig:CyclotronAndSpinFlipLineshape}a shows $\sigma^2 = \sigma_f^2 - \sigma_0^2$ vs.\ drive frequency.  The scale to the right in Fig.~\ref{fig:CyclotronAndSpinFlipLineshape}a is the average probability that the spin drive pulse makes a spin flip.  

Matching a 221 MHz drive so that the oscillating current divides equally between the two sides of an electrode (Fig.~\ref{fig:PbarMeasurementTrap}c) within a cryogenic vacuum enclosure is challenging, but is improved with a transmission line transformer for this measurement.  The drive applied still observably shifts $f_z$ as a function of spin drive power (Fig.~\ref{fig:MeasurementSequence}b), presumably because the average trapping potential is slightly modified, but less than for the p measurement. The shift from  the strongest drive in Fig.~\ref{fig:MeasurementSequence}b  is too small to contribute to the measurement uncertainty.    

The basic idea of the cyclotron frequency measurement is much the same as for the spin frequency.
The applied resonant drive is weak enough to cause no detectable growth in the average cyclotron radius and energy, but strong enough to increase the measured Allan variance, $\sigma_f^2$. The cyclotron lineshape (Fig.~\ref{fig:CyclotronAndSpinFlipLineshape}b) shows the expected sharp threshold at the trap cyclotron frequency, $f_+$.

For each of the drive frequencies represented in the cyclotron lineshape in Fig.~\ref{fig:CyclotronAndSpinFlipLineshape}b a cyclotron drive is applied continuously for 2 to 4 hours.  Deviations $\Delta_i$ between consecutive 80 s $f_z$ averages  
are characterized by an Allan variance, $\sigma_f^2$.  Then $\sigma_0^2$ (from below the threshold frequency) is subtracted to get $\sigma^2$.

No fits to expected resonance lineshapes are used for this measurement, but the
spin lineshape fits well to the Brownian motion lineshape \cite{BrownLineshape} expected for magnetic field fluctuations caused by thermal axial motion within a magnetic bottle gradient for a spin 1/2 system.  An axial temperature of 8 K is extracted from the fit, consistent with measurements using a magnetron method detailed in \cite{OneProtonSelfExcitedOscillator}.  With no expected lineshape yet available for the cyclotron resonance, we note that the cyclotron line  fits well to the expected spin lineshape but with an axial temperature of 4 K. A proper diffusion treatment of the way that a cyclotron drive moves population between cyclotron states is still needed.

A ratio of frequencies determines the magnetic moment in nuclear magnetons (Eq.~\ref{eq:ProtonMagneticMoment}).  The free space cyclotron frequency, $f_c = e B/(2 \pi m_p)$, is needed while trap eigenfrequencies $f_+$, $f_z$ and $f_{\_}$ are measured directly.  The Brown-Gabrielse invariance theorem, $f_c^2 = f_+^2 + f_z^2 + f_{\_}^2$ \cite{InvarianceTheorem} determines $f_c$ from the eigenfrequencies of an (unavoidably) imperfect Penning trap.

Applying Eq.~\ref{eq:ProtonMagneticMoment} gives the measured \pbar magnetic moment
\begin{equation}
\mu_{\pbar}/\mu_N = -2.792\,845 \pm 0.000\,012~~~~~[4.4~\rm{ppm}].
\label{eq:Result}
\end{equation}
The total uncertainty, with all known contributions detailed in Table~\ref{table:Uncertainties}, is 680 times smaller than obtained in an exotic atom measurement.  Frequency uncertainties are the half-widths of the sharp edges in the lineshapes, determined less precisely than for \mup because larger frequency steps were taken.  The magnetron linewidth uncertainty comes from the distribution of magnetron radii following sideband cooling done without and with simultaneous axial feedback cooling \cite{FeedbackCoolingPRL,OneProtonSelfExcitedOscillator}
for the spin and cyclotron cases.

\begin{table}[htbp!]
    \centering  
      \begin{tabular}{llr}

                \hline\hline

                Resonance & ~~~~~Source & ppm\\

                \hline\hline

                spin      & resonance frequency  & 2.7 \\

                \hline

                spin   &  magnetron broadening & 1.3\\
                \hline

                cyclotron~~~~~ & resonance frequency   & 3.2 \\

                \hline

                cyclotron & magnetron broadening & 0.7\\
                \hline \hline

                total & & 4.4 \\

                \hline\hline

            \end{tabular}

        \caption{Significant uncertainties in ppm.}\label{table:Uncertainties}
\end{table}

Comparing \mupbar to previously measured \mup gives  
\begin{eqnarray}
\mu_{\pbar}/\mu_p = &-1.000\,000~ &\pm 0.000\,005~~~~[5.0~\rm{ppm}]\\
\mu_{\pbar}/\mu_p = &-0.999\, 999\, 2 &\pm 0.000\,004\,4~~[4.4~\rm{ppm}],
\end{eqnarray}
consistent with the prediction of the CPT theorem.     
The first uses the \mup directly measured within the same trap electrodes \cite{ProtonMagneticMomentWithCorrection}.  The second uses the more precise \mup deduced indirectly from three measurements (not possible with \pbar) and two theoretical corrections \cite{MITProtonMoment,CODATA1998}.
  
A comparison of the \pbar and \p moments that is $10^3$ to $10^4$ times more precise seems feasible, to make a baryon CPT test with a precision approaching the $9 \times 10^{-11}$ comparison of the charge-to-mass ratios of \pbar and p \cite{FinalPbarMass}.  Individual spin flips must be resolved so quantum jump spectroscopy can be used to measure $f_s$, as done to measure the electron magnetic moment \cite{HarvardMagneticMoment2008}.  The jitter of $f_z$ described above is the challenge since in one measurement cycle this is not a lot smaller than the shift from a spin flip.  So far we can determine the spin state with a high probability in about 1 of 4 attempts but are hopeful that this efficiency can be improved to allow making a spin resonance in a reasonable time. The magnetic gradient used to detect an electron spin flip was small enough that flipping and detecting the spin could be done in the same trap.  To avoid the line broadening due to the large magnetic gradient needed to detect a \pbar spin state, spin flips must be driven in an adjacent trap with no magnetic gradient before being transferred to the trap used for spin state detection (as done with ions \cite{QuintDoublePenningTrap}). Measuring $f_c$, the second frequency needed to determine the \pbar magnetic moment, has already been demonstrated to better than $10^{-10}$ \cite{FinalPbarMass} in a trap with essentially no magnetic gradient.

In conclusion, a direct measurement of the \pbar magnetic moment to 4.4 ppm is made with a single \pbar suspended in a Penning trap, improving upon the value from exotic atom spectroscopy by a factor of 680.  The measured ratio of the \pbar and p magnetic moments is consistent with the value of -1 predicted by the CPT theorem to 5 ppm or better.    It seems possible in the future to compare the magnetic moments of \pbar and \p  more precisely, by a factor of $10^3$ or $10^4$ in addition to what is reported here.  

Thanks to CERN for the 5-MeV \pbar and some support for W.O, and to N.\ Jones for helpful comments.  This work was supported by the US NSF and AFOSR, and the Canadian NSERC, CRC, and CFI.

\newcommand{\Desktop}{\bibliography{d:/Jerry/Shared/Synchronize/ggrefs2012}}
\newcommand{\Laptop}{\bibliography{c:/Users/gabrielse/Jerry/Shared/Synchronize/ggrefs2012}}

%

\end{document}